\documentclass[11pt]{article}
\usepackage{deauthor}
\usepackage{amsmath}
\usepackage{booktabs}
\usepackage{caption}
\usepackage{comment}
\usepackage{epsfig}
\usepackage{graphicx}
\usepackage{subfig}
\usepackage{times}
\usepackage{url}
\usepackage{wrapfig}
\usepackage{xcolor}
\usepackage{xspace}
\usepackage{listings}
\usepackage{array}
\usepackage{multirow}
\usepackage{enumitem}
\usepackage{wrapfig}
\usepackage[hidelinks]{hyperref}
\usepackage[skins,breakable]{tcolorbox}
\usepackage[skip=3pt]{caption}
\setlist{nolistsep}
\usepackage[compact]{titlesec}
\newcommand{\para}[1]{\vspace{0.5mm}\noindent\textbf{#1}}

\newcommand{\exampleblock}[2]{
    \par \vspace{0.2em} 
    \noindent\textbf{Example #1:} \textit{#2}
    \vspace{0.2em} 
}

\begin{document}

\title{Towards Trustworthy and Cost-Efficient Data Integration: From Na\"ive RAG to Agentic RAG}

\author{
  \begin{tabular}{cc}
    Chuangtao Ma & Arijit Khan \\[2pt]
    \multirow{2}{*}{Aalborg University, Denmark} & Bowling Green State University, USA \\ 
    & Aalborg University, Denmark \\
    \texttt{chuma@cs.aau.dk} & \texttt{arijitk@bgsu.edu}
  \end{tabular}
}

\maketitle

\begin{abstract}
Large language models (LLMs) and AI agents have demonstrated strong potential for data integration in zero-shot and few-shot settings. However, they continue to face significant accuracy and cost challenges in enterprise environments due to a persistent knowledge gap. This paper envisions trustworthy, scalable, and cost-efficient integration through knowledge-grounded LLMs and agents operating within a retrieval-augmented generation (RAG) workflow. Here, \textit{trustworthiness} refers to evidence-grounded, verifiable reasoning, where integration decisions are transparently supported by retrieved knowledge, robust against hallucination, and consistent across tasks. We trace the evolution from classic RAG to GraphRAG and KG-RAG (knowledge graph-based RAG), highlighting how these paradigms bridge parametric and contextual knowledge. Building on this trajectory, we explore the shift toward Agentic RAG, where autonomous multi-agent systems adaptively plan, retrieve, refine, and reason for complex integration tasks. We examine optimization strategies for cost-efficient integration, addressing computational bottlenecks in large-scale enterprise settings. Finally, we outline open challenges and future directions toward building reliable, explainable, and scalable knowledge‑grounded integration systems.
\end{abstract}

\section{Introduction}
Data integration addresses heterogeneity by reconciling inconsistencies and conflicts across multiple sources to provide a unified view for data analytics and machine learning~\cite{AmalurRihan2023}, making it a fundamental task in modern data engineering and management. Integration is typically decomposed into subtasks such as schema matching, entity matching, entity resolution, column type annotation, and column property annotation, etc. However, semantic heterogeneity across sources poses substantial obstacles to establishing correspondences and resolving conflicts~\cite{Putrama2024heterogeneous}. In particular, domain‑specific abbreviations and acronyms in table, entity, and column names, combined with diverse naming conventions, create significant challenges for reliable integration.

\exampleblock{1}{\noindent
\noindent
In medical and healthcare data integration, systems must reconcile heterogeneous representations across sources. A schema matcher must recognize that abbreviations and acronyms denote patient identifiers, such as \texttt{MRN} (Medical Record Number) in an electronic health record (EHR) system and \texttt{Pat\_ID} in a laboratory information system. An entity matcher must determine that \texttt{J.~Smith, DOB 1985‑03‑12}'' in the EHR and \texttt{John A.~Smith, 03/12/85}'' in the lab system refer to the same individual. A column type annotation (CTA) system must annotate the \texttt{ICD10} code column as \texttt{Medical\_Code} rather than \texttt{Product\_Code}.
}

To address heterogeneity across data sources, data integration has been studied through rule‑based heuristics~\cite{Fan018}, supervised machine learning~\cite{Hai2023amalur}, and pre‑trained language models (PLMs)~\cite{TuFTWL0JG23}. Rule‑based methods rely heavily on domain expertise to design and maintain rules, while machine learning and PLM approaches demand large labeled datasets and incur substantial computational cost. Recent advances in LLMs and AI agents have driven a paradigm shift from supervised learning to in‑context learning~\cite{FreireFFKLPSSW25,NarayanCOR22}. LLMs now demonstrate competitive performance on core integration tasks including schema matching~\cite{ParciakVNPV25}, entity matching~\cite{PeetersSB25}, entity resolution~\cite{FuT0MKG25}, and column type annotation~\cite{euerLHF24} under in‑context learning and fine‑tuning settings.

Nevertheless, LLMs face systemic limitations that hinder their effectiveness for enterprise‑scale data integration. They suffer from both false negatives and false positives, reflecting two sides of the same knowledge gap. (1) Enterprise integration is inherently difficult: evolving schemas, heterogeneous sources, and domain‑specific conventions demand continuous adaptation. LLMs and Agents make matching decisions primarily on static parametric knowledge and lack access to domain‑specific and up‑to‑date contextual information, leading to missed matches (false negatives). (2) Severe class imbalance and hallucination tendencies undermine accuracy: datasets dominated by non‑matching pairs amplify inherent biases, while superficial lexical cues mislead models into spurious matches (false positives). Limited reasoning capabilities exacerbate both error types. (3) The substantial scale of enterprise tables and entities, coupled with diverse column types (e.g., abbreviations, numeric values), imposes prohibitive computational overhead due to the large number of comparisons required. As a result, LLM-based integration often suffers from performance degradation and high cost~\cite{MaCERAG4EM26,BodensohnBVSB25}. Beyond accuracy and efficiency, a critical missing dimension is \textit{trustworthiness}, defined as evidence-grounded, verifiable reasoning. Trustworthiness ensures that integration decisions are transparently supported by retrieved knowledge, robust against hallucination, and logically consistent across tasks. Unlike traditional evaluation focused solely on matching accuracy, it emphasizes validation and explainability, making integration reliable for enterprise deployment. These limitations arise from the prevailing paradigm in which LLMs act as matchmaker~\cite{SeedatS25}, generating outputs solely from internal knowledge. Without grounding in structured and contextual evidence, such out-of-the-box models fail to deliver robust integration in complex enterprise settings~\cite{KayaliWTD25}.

We argue that LLM-based data integration must shift from fine‑tuned and in‑context learning approaches to knowledge‑grounded paradigms that ground LLMs with precise external knowledge for complex reasoning. The central question motivating this work is: \emph{How can we ground LLMs and Agents with external knowledge to make data integration trustworthy, scalable, and cost‑efficient?} 
This paper outlines a vision for trustworthy and cost-efficient data integration from na\"ive RAG to agentic RAG.
We summarize our contributions below:
\begin{itemize}[leftmargin=*]
\item We present a systematic analysis of the challenges and knowledge gaps in LLM-based data integration, along with the technical limitations of knowledge-grounded approaches, tracing their evolution from na\"ive RAG to GraphRAG and KG‑RAG.
\item We propose a vision for trustworthy and cost-efficient data integration with agentic RAG, and design a roadmap and architecture that incorporate adaptive retrieval, batch processing, and graph-based memory.
\item We highlight key research gaps and opportunities for advancing trustworthy and cost-efficient agentic RAG-based data integration.
\end{itemize}
  
The remainder of this paper is organized as follows. In \S\ref{sec:knowledge_gap}, we examine the challenges of LLM-based data integration and the underlying knowledge gap that hinders trustworthiness. In \S\ref{sec:data_integration_with_rag}, we review emerging knowledge-grounded integration systems and highlight their technical limitations. Building on these insights, \S\ref{sec:towards_trustworthy_cost_efficient} outlines our vision and roadmap for trustworthy and cost-efficient agentic RAG-based integration. Finally, \S\ref{sec:research_gap_opportunities} summarizes the key research gaps and opportunities for advancing agentic RAG in data integration.

\section{Bridging the Knowledge Gap for LLM-based Data Integration} \label{sec:knowledge_gap}
This section examines the systemic challenges of LLM-based data integration and highlights the knowledge gap that underlies them, motivating the shift toward knowledge‑grounded approaches.
\subsection{Challenges of LLM for Data Integration}
\begin{wrapfigure}{R}{0.5\textwidth}
\centering
    \centering
    \includegraphics[width=1\linewidth]{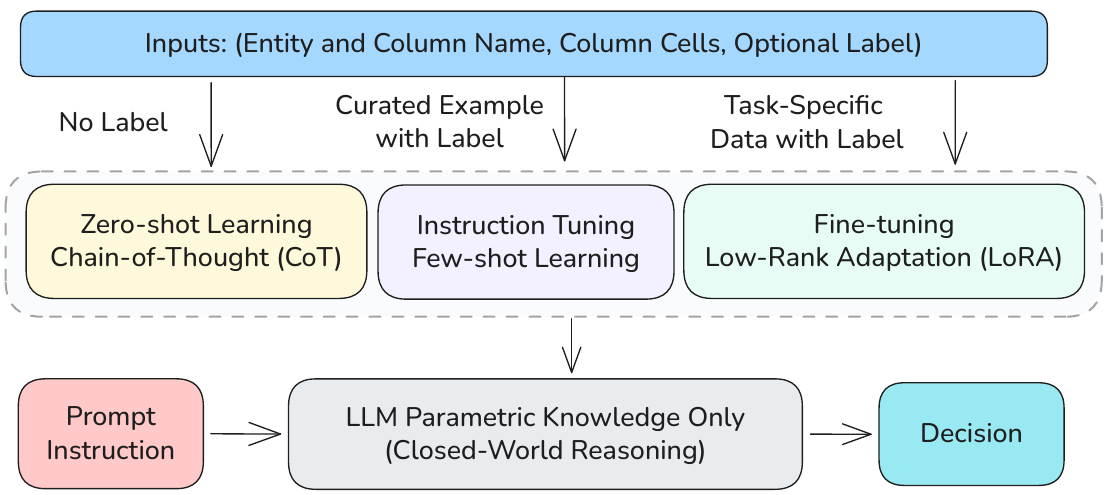}
    \caption{LLM for Data Integration.}
    \label{fig:llm4dataintegration}
\end{wrapfigure}

LLMs have been applied to schema matching, entity matching, and column type annotation using paradigms such as zero-shot learning~\cite{SeedatS25,Zhang2026Gather} with chain-of-thought (CoT) reasoning~\cite{Rohan25MultiRea}, few-shot learning with instruction tuning~\cite{MugeniLAM25}, and supervised fine-tuning (SFT) with low-rank adaptation (LoRA)~\cite{steiner2025fine}, as depicted in Figure~\ref{fig:llm4dataintegration}. 
Despite their strong reasoning capabilities, LLM-based integration suffers from three interrelated limitations, all rooted in reliance on closed-world parametric knowledge.

\para{Hallucination and Class Imbalance.} LLMs frequently generate incorrect correspondences when relying solely on internal parametric knowledge. In some cases, they miss true matches (false negatives) due to a lack of domain-specific or contextual knowledge. In other cases, they hallucinate spurious matches (false positives), misled by superficial lexical or syntactic cues~\cite{MaCERAG4EM26}. This dual vulnerability is amplified by extreme class imbalance: candidate pairs generated via Cartesian product are dominated by non-matching pairs, making true matches rare~\cite{Putrama2024heterogeneous}. As a result, LLMs struggle to learn the sparse target distribution, undermining accuracy.

\para{Dependence on Labeled Data.} While LLMs reduce the need for task-specific labeled data compared to traditional PLMs and machine learning approaches, fine-tuning, few-shot learning, and reward-based agent optimization still require curated examples with labeled data. For instance, fine-tuning even small models for entity matching demands thousands of labeled pairs~\cite{Hai2023amalur}. In practice, such labeled data is often unavailable in enterprise-scale data integration scenarios, limiting applicability.

\para{High Computational and Inference Costs.} LLM-based integration incurs prohibitive costs in fine-tuning, inference, and token consumption. Without external knowledge grounding, models must rely on parametric memory for complex reasoning, leading to long inference chains and quadratic scaling of API calls due to pairwise matching. Recent works show that fine-tuning Jellyfish-8B requires multiple GPUs and hours of training, while fine-tuning GPT-4o for entity matching costs tens to hundreds of dollars even with small datasets~\cite{MaCERAG4EM26,BodensohnBVSB25}. Such overheads make LLM-based integration impractical for large-scale enterprise data engineering.

Together, these challenges illustrate that LLMs, when treated as probabilistic “black box” matchmaker~\cite{SeedatS25}, fail to deliver trustworthy and cost-efficient integration. Their reliance on static parametric knowledge leaves them unable to adapt to evolving schemas, heterogeneous sources, and domain-specific conventions~\cite{KayaliWTD25}.

\subsection{Knowledge-Grounded LLMs and AI Agents for Trustworthy Data Integration}
The fundamental limitation of current LLM-based and agent-based integration paradigms is the absence of contextual knowledge for verifiable, fact-based reasoning. RAG addresses this gap by augmenting inference with factual evidence retrieved from external sources. Applied to schema and entity matching, RAG enables decisions grounded in curated knowledge bases (e.g., Wikidata, DBpedia, domain-specific KGs).

\para{Contextual Knowledge Mitigates Hallucination.} 
By anchoring inference with the retrieved evidence, RAG reduces hallucination and improves accuracy~\cite{MaKGRAG4SM25, MaCERAG4EM26}. For example, rather than misinterpreting “MRN” as “Machine Registration Number,” a knowledge-grounded system retrieves the relevant subgraph from healthcare KGs (e.g., UMLS), where ``MRN'' is explicitly defined as ``Medical Record Number''. This evidence guides the agent to the correct decision and provides a reasoning chain for verification.

\para{Reducing Labeled Data Requirements.} Knowledge grounding alleviates dependence on curated examples. By supplying explicit contextual evidence, RAG reduces the need for fine-tuning and few-shot demonstrations, enabling integration tasks to be performed without extensive labeled datasets.

\para{Lowering Computational Costs.} Grounded inference reduces token consumption and API calls by replacing long parametric reasoning chains with concise, evidence-based decisions. Batch retrieval and batch inference further optimize costs by reusing shared context across queries, avoiding redundant retrieval and inference.

This motivates a paradigm shift from purely parametric LLM-based integration toward knowledge-grounded approaches. By grounding inference with the retrieved contextual knowledge, RAG and its extensions (GraphRAG, KG-RAG, and Agentic RAG) provide a pathway to trustworthy, scalable, and cost-efficient data integration. 

\section{Data Integration with RAG, GraphRAG, and KG-RAG} \label{sec:data_integration_with_rag}
\begin{wrapfigure}{R}{0.65\textwidth}
\centering
    \centering
    \includegraphics[width=1\linewidth]{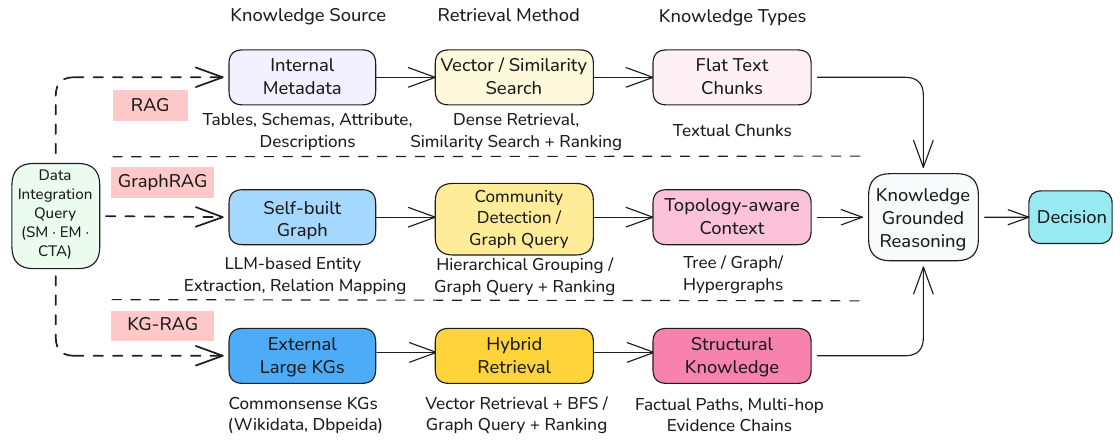}
    \caption{Data Integration with RAG, GraphRAG, and KG-RAG.}
    \label{fig:ragcomparision}
\end{wrapfigure}

This section traces the evolution of knowledge‑grounded data integration, beginning with vector‑based RAG, advancing through structure‑aware GraphRAG, extending to KG‑RAG with curated large‑scale knowledge graphs, and culminating in hybrid semantic retrieval. 
As depicted in Figure~\ref{fig:ragcomparision}, these advances illustrate the progression from flat textual evidence to structured, multi‑hop, and adaptive knowledge grounding.

\subsection{Data Integration with RAG}
Retrieval-Augmented Generation (RAG) provides a principled approach to data integration by reducing the risk of hallucinations in large language model (LLM)-based systems. Instead of relying solely on the generative capabilities of the LLM, RAG grounds both the model and its agentic workflow in contextual knowledge retrieved from authoritative sources. In practice, this contextual knowledge is often drawn from internal metadata such as database design documents. These documents capture the logical structure of the system, including data tables, attributes, and textual descriptions that explain how tables and attributes are connected.

\exampleblock{2}{\noindent
Consider the database design document of the \texttt{e-MedSolution} data model~\cite{ma2020knowledge}. Each table and attribute is accompanied by a textual description that clarifies its meaning and relationships. For example:
\begin{itemize}
    \item \texttt{MRN}: The hospital patient identifier assigned upon admission that acts as a foreign key in the \texttt{hun\_patient} table, linking each record to the corresponding patient.
    \item \texttt{person\_id}: A unique identifier for each patient, used to track individuals who are at risk and to record their clinical observations in the source system.
\end{itemize}
By retrieving and grounding such metadata, RAG ensures that integration tasks such as schema matching are guided by the actual design logic of the database rather than by potentially unreliable model inferences.
}

Early work demonstrates that RAG mitigates hallucinations in LLM-based integration without requiring task-specific labeled data, grounding inference in contextual evidence rather than relying solely on parametric memory. For instance, ReMatch transforms schema attribute metadata into structured documents and retrieves candidate attributes via dense search, with top-ranked results semantically ranked by an LLM, thereby eliminating the need for labeled training data in schema matching~\cite{sheetrit2024rematch}. 
While na\"{\i}ve RAG approaches reduce hallucination and bypass training requirements, they remain shallow for trustworthy integration. Their reliance on retrieved flat textual contexts limits the ability of LLMs to leverage structural knowledge, constraining robust grounding in complex data integration scenarios~\cite{MaKGRAG4SM25, MaCERAG4EM26, wei2024racoon}.

\subsection{Data Integration with GraphRAG}
GraphRAG extends RAG by grounding LLMs in structured, topology-aware knowledge such as graph contexts~\cite{traeger2026ract}, context trees, and hypergraphs~\cite{Chen2026ConstruM}, rather than relying on flat textual chunks. This structural knowledge is typically built from task-specific metadata through LLM-based entity extraction and relationship mapping. The resulting graph of interconnected nodes is organized via community-based hierarchical clustering~\cite{WangFZLM26} and then summarized by LLMs to provide richer context for knowledge augmentation.

Traditional similarity-based approaches often fail in multi-table schema matching because embeddings across different contexts exhibit low similarity. To address this, RACT~\cite{traeger2026ract} augments column representations with relational schema graphs and inter-table paths, replacing unstructured text retrieval with structured relational evidence. ConstruM~\cite{Chen2026ConstruM} advances this further by introducing multi-level context retrieval: lightweight context trees capture local structure, while global similarity hypergraphs provide broader relational grounding. Together, these methods enable more accurate candidate grouping and disambiguation.

Despite these advances, current RAG and GraphRAG frameworks remain constrained by limited knowledge coverage. Self-constructed graphs often capture only basic schema metadata such as table names, entities, attributes, key constraints, and short textual descriptions, while deeper relationships and factual connections across data sources are missing. Moreover, graph construction and indexing are computationally expensive, with clustering and summary overheads growing linearly with corpus size. In enterprise scenarios, internal metadata and design documents are frequently unavailable~\cite{BodensohnBVSB25}, leaving hallucinations unresolved when grounding lacks sufficient structural context.

\subsection{Data Integration with KG-RAG}
To address the limitations of RAG and GraphRAG, recent work has shifted toward KG-RAG-based data integration~\cite{wei2024racoon, MaKGRAG4SM25, Jeon2025SM}. KG-RAG grounds LLMs in richer contextual knowledge from external knowledge bases, enabling multi-hop reasoning across curated graphs. Unlike GraphRAG, which augments LLMs with small, self-constructed graphs derived from textual metadata, KG-RAG retrieves relevant subgraphs from large-scale knowledge graphs, providing broader factual coverage and stronger semantic grounding.

\exampleblock{3}{\noindent
Consider schema matching in clinical databases. Using only metadata, an LLM incorrectly matches \texttt{measurement\_time} in \texttt{measurement} table to \texttt{perfac\_date} in \texttt{hun\_case\_c\_i\_interventions} table, due to superficial textual similarity. In contrast, KG-RAG~\cite{MaKGRAG4SM25} retrieves an external subgraph containing multi-hop evidence from Wikidata: “beneficiary (Q2596417) $\rightarrow$ subclass of (Q21514624) $\rightarrow$ customer (Q852835) $\rightarrow$ subclass of (Q21514624) $\rightarrow$ patient (Q181600).” This retrieved structured evidence chain in KG-RAG shows that the beneficiary is a superclass of patient, which helps LLMs to distinguish clinical focus \texttt{measurement} from administrative and operational focus \texttt{intervention}.  By leveraging curated external KG, KG-RAG mitigates hallucinations and improves integration accuracy in complex, multi-table scenarios.
}

KG-RAG has been applied to key integration tasks such as schema matching~\cite{MaKGRAG4SM25}, entity matching~\cite{MaCERAG4EM26}, and column type annotation~\cite{wei2024racoon} by leveraging external factual knowledge graphs. For example, RACOON~\cite{wei2024racoon} improves column type annotation in domain-specific queries by augmenting LLMs with structural factual knowledge retrieved from Wikidata via a KG Linker, thereby reducing factual inaccuracies. KG‑RAG4SM~\cite{MaKGRAG4SM25} integrates vector-based retrieval, graph traversal, and ranking-based refinement in a hybrid pipeline. This approach retrieves and prunes subgraphs from large commonsense KGs, enabling LLMs to resolve complex semantic conflicts in schema matching. For entity matching, CE‑RAG4EM~\cite{MaCERAG4EM26} combines dense retrieval with graph-based expansion and traversal to extract structural evidence from Wikidata.

The use of large-scale external KGs provides richer factual grounding than self-built graphs, supporting deeper evidence chains for complex integration scenarios, particularly when internal metadata or background knowledge is unavailable. However, retrieving and exploring relevant subgraphs from large external KGs remains costly and time-consuming~\cite{wei2024racoon, MaCERAG4EM26}. Moreover, current RAG and GraphRAG systems often rely on uniform retrieval strategies across queries, regardless of complexity or ambiguity. This lack of adaptivity introduces unnecessary overhead and latency, limiting scalability in large-scale data integration.

\section{Advancing Data Integration with Agentic RAG  } \label{sec:towards_trustworthy_cost_efficient}

In this section, we outline the vision and design a roadmap for building trustworthy and cost-efficient data integration with agentic RAG.

\subsection{Roadmap and Vision}

Despite the advances of the static RAG, GraphRAG, and KG-RAG pipelines in data integration, they suffer from fundamental scalability and efficiency bottlenecks rooted in their static and per-query retrieval design. 
Agentic RAG~\cite{zhou2025towards} aims to address the above limitation of static RAG by dynamically retrieving the contextual knowledge on demand and adaptively selecting the optimal retrieval granularity for knowledge grounding. 

To address the above limitations of data integration with na\"ive RAG, GraphRAG, and KG-RAG, we design a roadmap and vision for building trustworthy and cost-efficient data integration with agentic RAG. 

\begin{wrapfigure}{r}{0.5\textwidth}
\centering
    \centering
    \includegraphics[width=1\linewidth]{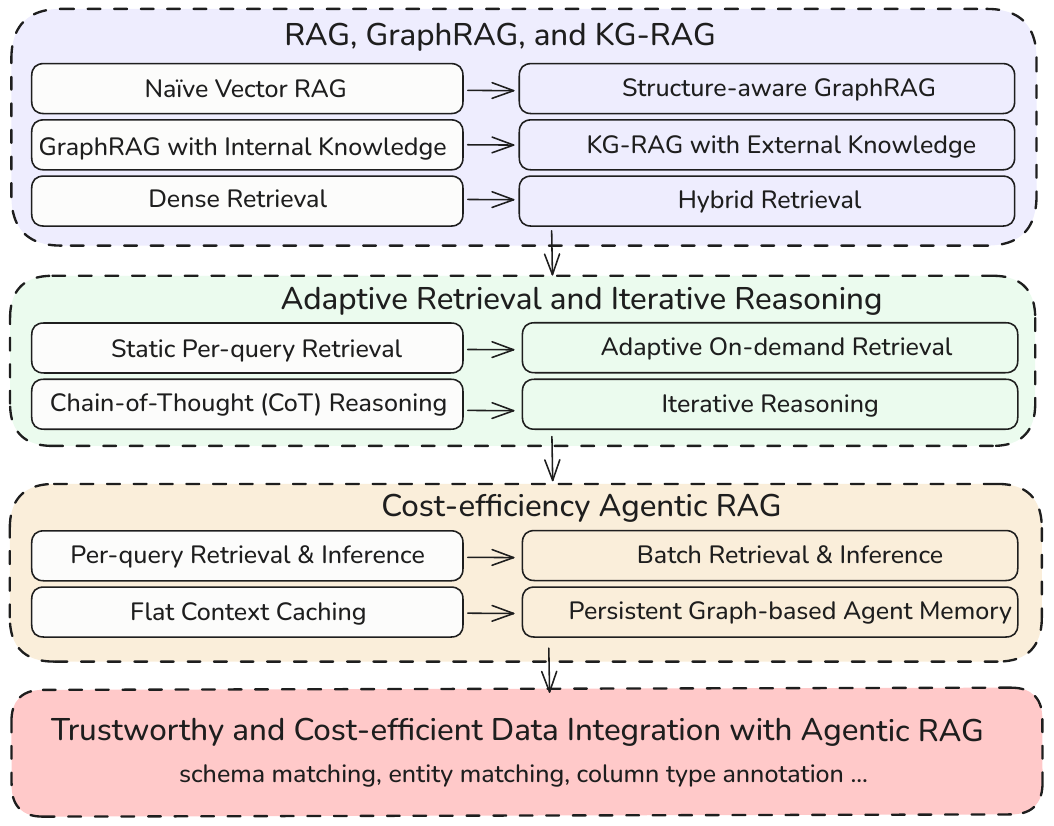}
    \caption{Vision Towards Trustworthy and Cost-Efficient Data Integration with Agentic RAG.}
    \label{fig:paradigm_shift}
\end{wrapfigure}
As shown in Figure~\ref{fig:paradigm_shift}, it presents a vision and roadmap of next-generation trustworthy and cost-efficient data integration with agentic RAG, illustrating the key components and paradigm shifts in building trustworthy and cost-efficient data integration with Agentic RAG.
First, architectural advances in different RAG directly address hallucination and trustworthiness gaps by grounding LLMs with precise evidence chains for complex reasoning. Second, adaptive retrieval and iterative reasoning enable systems to invoke retrieval only when necessary and to refine decisions across retrieved evidence, rather than relying on single-pass CoT inference. Third, efficiency-oriented strategies preserve trustworthiness while reducing cost by exploiting shared contexts across queries and eliminating retrieval cold-starts through graph-based persistent agentic memory.

\subsection{Adaptive Retrieval and Iterative Reasoning}
\begin{wrapfigure}{R}{0.5\textwidth}
\centering
    \centering
    \includegraphics[width=0.95\linewidth]{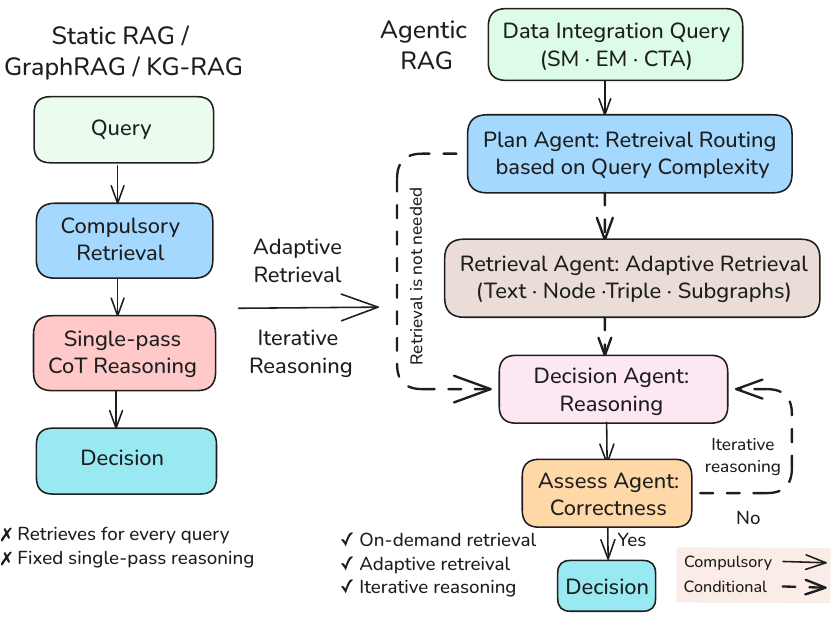}
    \caption{Toward Agentic RAG-based Data Integration.}
    \label{fig:agenticrag}
\end{wrapfigure}

Recent advances have witnessed the strengths of adaptive retrieval and iterative reasoning in Agentic RAG enabled by multi-agent collaboration, which motivates a vision toward agentic RAG-based data integration.
Figure~\ref{fig:agenticrag} outlines a paradigm shift and a vision toward agentic RAG-based data integration, where multi-agent collaboration enables adaptive retrieval and iterative reasoning to overcome these limitations. 
To make this actionable, we demonstrate how an agentic RAG with adaptive retrieval, iterative reasoning, and task planning mitigates the hallucination and cost issues inherent in na\"ive LLM and RAG-based data integration.

\para{From Static Retrieval to Adaptive Retrieval.}
In standard RAG, GraphRAG, and KG-RAG, retrieval is executed for every query, even when the model’s parametric knowledge is sufficient to resolve the task. Recent advances in Agentic RAG~\cite{Salve25MAgentRAG, Du26ARAG}, Graph‑RAG~\cite{LiS26coutRAG, Capozzi2026AgentGraphRAG}, and KG‑RAG~\cite{LelongEB25} shift away from static retrieval pipelines toward dynamic and adaptive retrieval. For instance, A‑RAG~\cite{Du26ARAG} selectively determines whether retrieval is necessary, thereby leveraging pre‑trained parametric knowledge while adapting retrieval behavior to task complexity.
Data integration typically involves a mix of straightforward queries with clear lexical similarity and a smaller set of ambiguous cases requiring multi‑hop inference and external knowledge grounding. In such scenarios, retrieval is often only needed for the latter, yet static RAG systems unconditionally perform retrieval for every query, incurring unnecessary cost at scale. 
Agentic RAG‑based integration~\cite{AlthafMMTC25} addresses this inefficiency by orchestrating dynamic retrieval through autonomous agents. These agents plan retrieval on demand, deciding whether retrieval is required, iteratively retrieving and reasoning when necessary, and assessing the sufficiency of contextual knowledge based on the complexity and uncertainty of each query.
To address the challenges of data heterogeneity in financial decision-making, Agentic GraphRAG~\cite{Capozzi2026AgentGraphRAG} integrates a data ingestion and an entity resolution pipeline into a collaborative agentic GraphRAG framework where a zero-shot intent routing and agent reflection loop are designed to identify the candidate entities and classify queries with disambiguated entities for adaptive graph exploration.

\para{From Chain-of-Thought Reasoning to Iterative Reasoning.}
Standard RAG and GraphRAG ground LLM inference through a single‑pass chain‑of‑ thought (CoT), executing fixed reasoning in one call. To overcome this limitation, recent work has advanced iterative retrieval and reasoning, marking a shift from static CoT to agent‑based iterative reasoning. For example, MCTS‑RAG~\cite{HuZZC25} integrates structured reasoning with adaptive retrieval, dynamically refining reasoning paths via Monte Carlo tree search. MA‑RAG~\cite{Nguyen25MARAG} extends this approach with multi‑agent orchestration, enabling agents to collaborate through iterative retrieval and intermediate reasoning with CoT prompting.
In data integration, ambiguous matching cases often demand multi‑hop iterative reasoning rather than a single‑pass inference. SMoG~\cite{Jeon2025SM} adapts this paradigm by iteratively uncovering explicit evidence through entity propagation, chain aggregation, and reasoning to identify optimal paths for schema matching. Multi‑Agent RAG~\cite{AlthafMMTC25} extends agentic RAG to entity resolution, coordinating specialized agents for task decomposition across linkage identification, record clustering, grouping, and relocation detection.

These advances define the frontier of agentic RAG for data integration: dynamic retrieval of evidence on demand, adaptive retrieval based on query complexity, and reasoning iteratively across it, rather than relying on static single‑pass CoT inference.

\subsection{Cost-Efficiency Strategies}
Although Agentic RAG reduces retrieval costs relative to static RAG and GraphRAG by adaptively retrieving contextual knowledge on demand, it still incurs substantial token consumption due to multi‑agent interactions and iterative reasoning~\cite{DangQLFXSCYCTXH25}. At large-scale data integration, the overall cost remains high, as integration tasks such as schema matching, entity matching, and column type annotation often involve thousands to millions of comparisons. In static RAG and KG‑RAG pipelines for schema matching~\cite{MaKGRAG4SM25}, each candidate pair triggers an independent retrieval from a large external knowledge base followed by a separate LLM inference call. These retrieval cost challenges highlight the need to move beyond vanilla RAG and static GraphRAG toward cost‑efficient RAG paradigms, which introduce new cost optimization strategies on retrieval and inference.

\para{From Per-Query Execution to Batch Processing.}
To reduce LLM costs and leverage shared structures in data integration, recent studies have explored batch prompting~\cite{JiBatchP25} and batch processing for entity resolution using in-context learning~\cite{FanHFC00024} and in-context clustering~\cite{FuT0MKG25}. 
BATCHER~\cite{FanHFC00024} and LLM-CER \cite{FuT0MKG25} achieve significant API cost savings over standard prompting by grouping multiple pairwise queries into a single batch with shared demonstrations. 
OBP~\cite{JiBatchP25} optimizes batch prompting by adaptively grouping questions and selecting demonstrations, yielding additional reductions in cost.
\begin{wrapfigure}{R}{0.45\textwidth}
\centering
    \centering
    \includegraphics[width=1\linewidth]{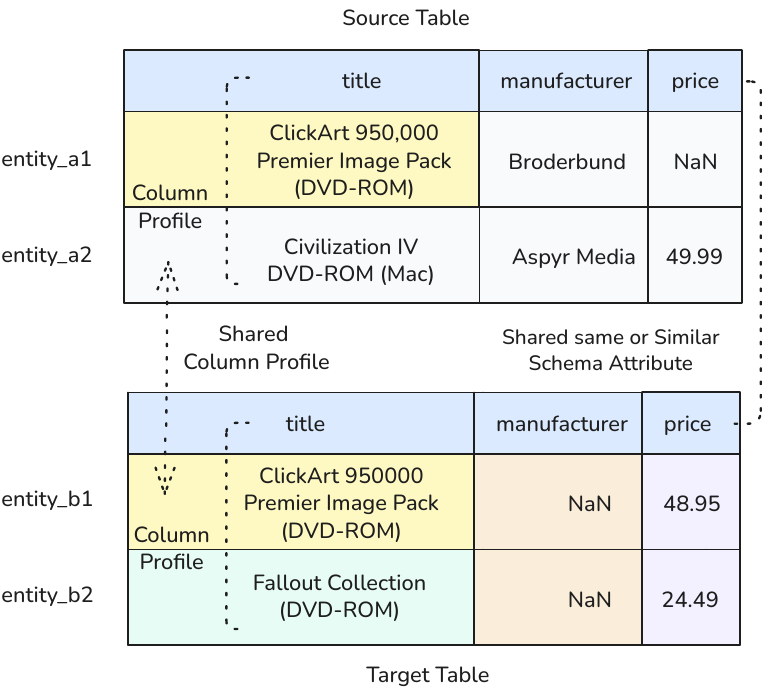}
    \caption{Structural Sharing in Data Integration.}
    \label{fig:structure_sharing}
\end{wrapfigure}
In RAG-based integration, na\"{\i}ve RAG and GraphRAG typically execute retrieval and inference independently for each query, ignoring structural similarities across tasks such as schema matching and column type annotation. 
As shown in Figure~\ref{fig:structure_sharing}, these shared structures include: \textbf{(1)} overlapping or similar attribute names between source and target tables, and \textbf{(2)} rows within the same column that share similar neighborhood context and column profiles.
Building on this insight, CE-RAG4EM~\cite{MaCERAG4EM26} introduces a blocking-based batch framework that retrieves contextual knowledge and inference once for a group of queries via batch retrieval and prompting, further reducing token consumption. 
Experimental results show that this strategy maintains or improves matching quality while significantly lowering cost compared to state-of-the-art baselines.
Overall, exploiting shared structures across tables and columns provides a natural foundation for batch processing in RAG-based data integration, enabling cost optimization without sacrificing performance.

\para{From Context Caching to Graph-Based Agentic Memory.}
Beyond batch processing, another avenue for cost optimization in RAG-based data integration is context caching and graph-based agentic memory. Early work~\cite{Jin2025RAGCache} explored simple caching strategies, reusing previously computed key-value (KV) stores or retrieved text chunks for similar or historical queries, thereby reducing redundant LLM prefill computation for shared prefixes. Building on this, CACHE-CRAFT~\cite{AgarwalSMMGSKYS25} introduced a cache management system that identifies reusable chunks and organizes them for effective reuse without compromising quality. However, these approaches rely on flat and unstructured caches that ignore structural relationships across tasks and domains. As a result, their effectiveness is limited to highly similar queries and degrades significantly when applied across domains. Moreover, cached contexts may not persist throughout RAG due to expiration and override.

To overcome these limitations, Agentic RAG~\cite{Lin2025CacheARAG} and GraphRAG~\cite{wu2026memgraphrag} introduce a persistent graph-based memory mechanism that enables memory search directly from the graph rather than re-embedding or re-querying external knowledge bases. 
This structured memory supports complex multi-hop reasoning and provides stable knowledge grounding for LLMs. 
Recent studies highlight a clear transition from cache-based optimization~\cite{AgarwalSMMGSKYS25} to graph-based agentic memory optimization~\cite{wu2026memgraphrag}. 
For example, MAGMA~\cite{Jiang2026MAGMA} introduces multi-graph agentic memory by organizing memory items across semantic, temporal, causal, and entity graphs, and formulating retrieval as policy-guided traversal over these relational views. 
MemGraphRAG~\cite{wu2026memgraphrag} further integrates memory-guided bridging for graph and memory co-evolution with memory-guided online retrieval. 
This enables query-adaptive selection and structured context construction, retrieving relevant evidence via multi-layer memory filtering and community-based search from the global context graph.

Motivated by this, Multi‑Agent RAG~\cite{AlthafMMTC25} integrates a global and persistent memory into multi-agent RAG for entity resolution via LangGraph, allowing asynchronous message passing and context sharing, which further enables efficient multi-agent collaboration without duplicate retrieval and computation.
CAIDA~\cite{Rand2026IntDataI} introduces a hybrid routing and retrieval agent for data integration that combines retrieval caches and long-term memory to balance knowledge grounding and latency.
Persistent graph-based memory thus provides direct knowledge grounding for LLMs in complex reasoning without incurring substantial retrieval or inference costs~\cite{AlthafMMTC25}, effectively addressing latency and scalability challenges in agentic RAG systems for large-scale data integration.

\para{From Linear Agent Call to Multi-Agent Orchestration.}
In agentic RAG with a single agent, subtasks are often executed sequentially, with each step awaiting the completion of the previous one. This linear workflow introduces significant latency~\cite{QiangWT24}. A \textit{multi-agent RAG framework} can overcome such latency and inefficiency in data integration. By distributing responsibilities across specialized agents, the framework enables parallelism and adaptive coordination, thereby reducing latency while maintaining effectiveness. Through adaptive retrieval, iterative reasoning, and task planning, multi-agent orchestration mitigates hallucination and lowers computational cost compared to naïve LLM- and RAG-based integration.
This orchestration reduces latency compared to linear single-LLM workflows, while maintaining lower retrieval costs than naïve RAG pipelines.

\subsection{System Architecture}
\begin{wrapfigure}{R}{0.6\textwidth}
\centering
    \includegraphics[width=1\linewidth]{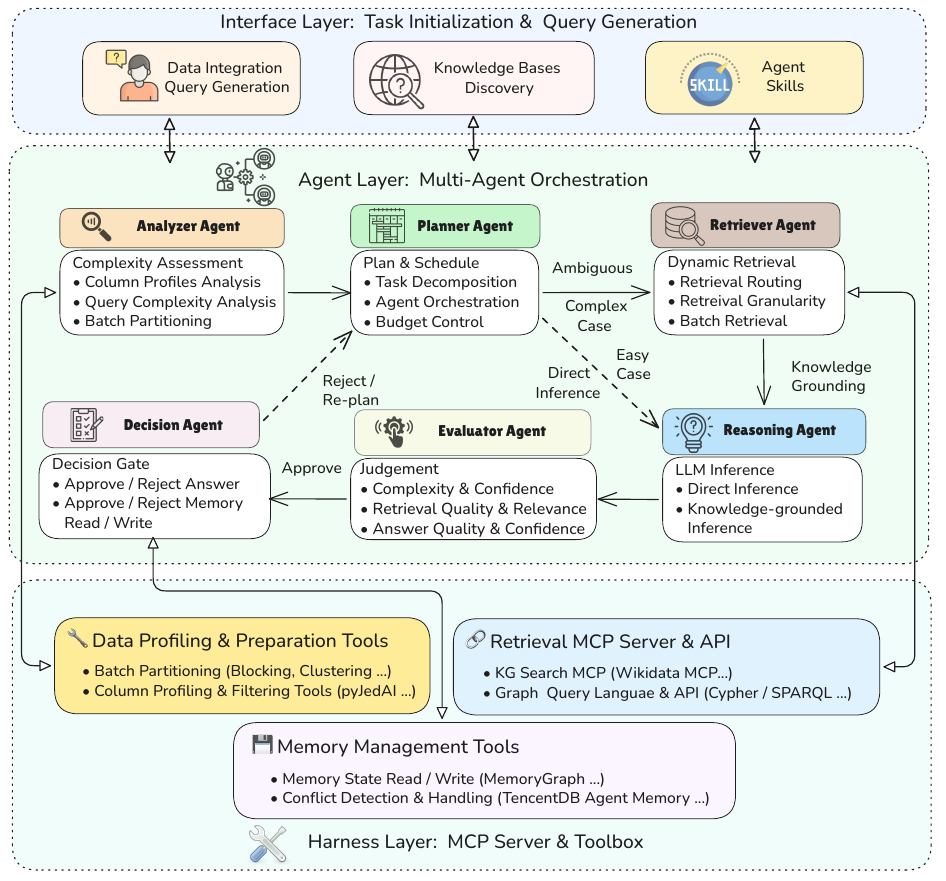}
    \caption{System Architecture of Multi-Agent RAG for Trustworthy and Cost-Efficient Data Integration.}
    \label{fig:system_architecture}
\end{wrapfigure}

Based on the above paradigm shift, we design the system architecture for trustworthy and cost-efficient data integration with agentic RAG.
As shown in Figure~\ref{fig:system_architecture}, our vision system is organized into three layers: \textsl{interface layer, agent layer, and harness layer}.
Each of the layers and components is tightly coupled to support the efficient retrieval and knowledge grounding over na\"ive-RAG-based data integration. 

The \textsl{interface layer} provides a user interface for the user to initialize the data integration task with query generation, knowledge base discovery, and configure the agent skills.
Then the generated query and discovered KGs, as well as the configured agent skills, are passed to the agent layer and the harness layer for task execution.

The \textsl{agent layer} consists of six key agents that are coordinated for task execution with the toolbox.

\textbf{Analyzer Agent:} It first analyzes the complexity of data integration queries using schema and column profiles (such as table names, column types, attribute diversity, and cell values). It then evaluates the query complexity level via LLM-as-judge, distinguishing easy cases with high confidence from ambiguous cases with lower confidence. This evaluation guides the \textsl{Planner Agent} in creating plans for task composition and agent scheduling. Finally, it dynamically groups queries into batches by invoking external blocking and clustering tools for batch processing.

\textbf{Planner Agent:} It creates a plan for each query by decomposing the overall integration task into an ordered sequence of sub-tasks based on the query’s complexity level. For each sub-task, it dynamically initializes and invokes agents equipped with the required skills and tools. In particular, it schedules retrieval routing and calls the \textsl{Retriever Agent} to access external knowledge at varying levels of granularity (entity, triple, and multi-hop subgraph) for ambiguous cases, while directly invoking the \textsl{Decision Agent} to produce answers for easy cases through direct inference. It also adjusts batch size and cost budget for each sub-task to jointly optimize matching accuracy and cost. Finally, it recreates plans for failed cases using feedback from the \textsl{Decision Agent} and \textsl{Evaluator Agent}.

\textbf{Retriever Agent:} It executes the retrieval plan generated by the \textsl{Planner Agent} by dynamically invoking MCP servers and the toolbox to adaptively access contextual knowledge across knowledge bases at varying levels of granularity (entity, triple, and multi-hop subgraph). It then engages the \textsl{Evaluator Agent} to assess the relevance and sufficiency of the retrieved knowledge for grounding. Finally, it appends the contextual knowledge to the decision prompt and calls the \textsl{Decision Agent} for knowledge-grounded inference.

\textbf{Reasoning Agent:} It generates outputs for a given query based on direct or knowledge-grounded inference, guided by the plan from the \textsl{Planner Agent} and contextual knowledge retrieved by the \textsl{Retriever Agent}.

\textbf{Decision Agent:} It evaluates whether the output answer is acceptable, manages memory operations based on feedback from the \textsl{Evaluator Agent}, and resolves conflicts between retrieved knowledge and memory by reinvoking the \textsl{Planner Agent} to generate a new plan for iterative retrieval and reasoning.

\textbf{Evaluator Agent:} It assesses each query by invoking LLM-as-judge to evaluate the confidence score of the \textsl{Decision Agent} output and the quality and relevance of contextual knowledge retrieved by the \textsl{Retriever Agent} for knowledge grounding. It monitors decision and retrieval quality, approving or rejecting outputs to ensure that only verifiable, faithful, and trustworthy results are committed. Rejected cases are returned to the \textsl{Decision Agent} and the pipeline is reinvoked for iterative retrieval and reasoning with minimal human supervision.

The \textsl{Harness layer} consists of a set of toolboxes including data profiling and preparation tools, retrieval MCP server and API, and memory management tools.
These toolboxes are designed to assist agents to exectue a spefic taks by dynamically calling different tools via MCP protocol and API on-demand.

Overall, these design choices enable the LLM multi-agent system to improve performance iteratively without significantly increasing cost through inference and knowledge grounding supported by dynamic, adaptive retrieval. Furthermore, integrated dynamic retrieval and persistent memory mechanisms ensure scalability and efficiency in large-scale data integration while maintaining trustworthiness.

\section{Research Gaps and Opportunities} \label{sec:research_gap_opportunities}
We outline below the research gaps and opportunities for advancing the vision of trustworthy and cost-efficient data integration with agentic RAG.

\para{Context and Memory Knowledge Conflict.} Graph-based agentic memory systems persist and reuse evidence across queries, significantly reducing retrieval costs in stable knowledge domains. However, enterprise data integration tasks such as schema and entity matching, as well as external knowledge bases, evolve over time: fields and attributes may change, and factual knowledge is periodically updated. Conflicts arise when the agentic RAG system relies on outdated memory while simultaneously retrieving updated context, leading to incorrect results due to the default prioritization of memory. Future research should investigate dynamic priority policies and incremental memory updating to address schema drift, entity evolution, and knowledge base updates.

\para{Batch Retrieval Noise and Adaptive Retrieval Estimation.} In RAG-based data integration, batch retrieval allows shared context across queries, but it can introduce retrieval noise for individual queries. Agentic RAG mitigates this by adaptively retrieving contextual knowledge only when parametric knowledge is insufficient, thereby reducing both cost and exposure to noise. However, batch and adaptive retrieval may fail in scenarios with high query diversity or highly imbalanced ground-truth distributions, since the absence of a default retrieval policy can prevent necessary evidence from being retrieved. Future work should investigate metrics to formally define and quantify retrieval noise, assess its impact in batch retrieval, and develop unified adaptive retrieval policies that estimate per-pair confidence and trigger retrieval based on uncertainty levels.

\para{Parameter Tuning and Agent Initialization.} Agentic RAG-based data integration systems with batch processing demand careful parameter configuration, including retrieval granularity, context length, batch size, and agent models. These must be tuned before deployment, yet labeled ground truth is typically unavailable, complicating fine-tuning and reward-based training. For instance, executing a new integration task over unseen product catalogs requires parameter selection and agent initialization without access to labels, creating a cold-start challenge. Addressing this issue calls for unsupervised and semi-supervised strategies such as uncertainty-based thresholds for adaptive retrieval, clustering-based initialization to group similar queries, Bayesian optimization guided by proxy metrics like consistency or stability, and meta-learning to transfer knowledge from related tasks. Additionally, high-capacity LLMs can serve as judges over limited datasets, providing weak supervision signals that enable semi-supervised initialization with reliable guidance.

\para{Autonomous Data Integration with Agentic Tool Orchestrations.} Current agentic RAG-based data integration systems~\cite{AlthafMMTC25, Rand2026IntDataI} focus on adaptive retrieval and dynamic inference through multi-agent coordination but fall short of full autonomy. Emerging frameworks~\cite{zhou2025towards} highlight the potential for autonomous data integration loops with minimal human oversight. In these systems, agents independently discover external knowledge sources, generate queries, construct column profiles via external toolboxes, adaptively retrieve evidence using protocols such as MCP (Model Context Protocol), and continuously resolve knowledge conflicts while balancing accuracy and cost. A central enabler is the \textit{tool calling graph}~\cite{LiuPCBZ0CWYD25}, which formalizes agent-tool orchestration: nodes represent specialized agents or external tools, and edges capture invocation dependencies and information flow. By leveraging this graph, autonomous agentic RAG systems can coordinate analyzer, planner, retriever, decision, and evaluator agents with their respective tools, ensuring robust execution of complex integration tasks. This opens opportunities for developing agent-compatible toolboxes and principled orchestration strategies that advance trustworthy and cost-efficient autonomous data integration.

\para{Cross-Task Enterprise Data Integration Benchmark and Metric.} Existing RAG-based data integration systems are evaluated on public benchmark datasets, which often overestimate performance in enterprise scenarios~\cite{BodensohnBVSB25}. Current benchmarks and metrics are largely designed for supervised machine learning approaches, emphasizing matching accuracy while neglecting dimensions critical to agentic RAG systems. In particular, they fail to assess \textit{trustworthiness}, which refers to the reliability of multi-hop reasoning, the grounding of decisions in verifiable evidence, and robustness against hallucination. To advance evaluation, future work should develop an enterprise-scale benchmark that unifies multiple integration tasks, including schema matching, entity resolution, and column type annotation, while incorporating ambiguous pairs to stress-test reasoning. Complementary metrics should measure both cross-task accuracy and trustworthiness, capturing whether integration decisions are not only correct but also transparently supported by evidence and logically consistent across tasks.

\section{Conclusion}
This paper examined the evolution of RAG-based data integration, tracing the progression from na\"{\i}ve RAG with dense retrieval, to KG-RAG with hybrid retrieval, to agentic RAG with adaptive retrieval, and finally to cost-efficient agentic RAG with batch processing. Building on this trajectory, we outlined a vision for trustworthy and cost-efficient data integration enabled by agentic RAG. We further proposed a multi-agent workflow that coordinates specialized agents through tool orchestration and identified key research gaps, including trustworthiness evaluation, adaptive retrieval policies, memory management, and benchmark design that must be addressed to realize fully autonomous, scalable, and reliable agentic RAG-based data integration systems.

\section*{Acknowledgment}
Chuangtao Ma and Arijit Khan acknowledge support from the Novo Nordisk Foundation grant NNF22OC0072415.

\bibliographystyle{IEEEtran} 
\bibliography{sample-base}

\end{document}